\documentclass[%
 reprint,
 amsmath,amssymb,
 aps,
]{revtex4-1}

\usepackage{graphicx}

\begin{document}

\preprint{APS/123-QED}

\title{Particle motion associated with wave function density gradients}

\author{Jan Klaers}
\email{j.klaers@utwente.nl}

\author{Violetta Sharoglazova}%
\author{Chris Toebes}%
\affiliation{
 Adaptive Quantum Optics (AQO), MESA$^+$ Institute of Nanotechnology, University of Twente, 7522 NB Enschede, Netherlands
}

\begin{abstract}
We study the quantum mechanical motion of massive particles in a system of two coupled waveguide potentials, where the population transfer between the waveguides effectively acts as a clock and allows particle velocities to be determined. Application of this scheme to evanescent phenomena at a reflective step potential reveals an energy-velocity relationship for classically forbidden motion. Regions of gain and loss, as described by imaginary potentials, are shown to speed up the motion of particles. We argue that phase and density gradients in quantum mechanical wave functions play complementary roles in indicating the speed of particles.
\end{abstract}

\maketitle

Quantum mechanics knows a variety of velocity definitions, such as phase velocity, group velocity, signal velocity \cite{sommerfeld1914fortpflanzung, brillouin1914fortpflanzung, ranfagni1991role}, velocities derived from the probability density flow or simply from the momentum operator. Particle velocities play a decisive role in the phenomenon of superfluidity \cite{Landau41,Anderson66,Legget99}, for example. For a superfluid described by the wave function $\psi(\mathbf{x},t)=\sqrt{n(\mathbf{x},t)}\exp(iS(\mathbf{x},t))$, where $n(\mathbf{x},t)$ denotes the density and $S(\mathbf{x},t)$ the phase, the superfluid velocity is defined by
\begin{equation}
\label{eq:velocity}
\mathbf{v}_s(\mathbf{x},t)=\frac{\hbar}{m}\nabla S(\mathbf{x},t) \;.
\end{equation}
This expression ascribes the velocity exclusively to phase gradients -- but not to density or amplitude gradients. The study of velocities in quantum mechanics presents several difficulties. Wave functions can contain superpositions of waves traveling in opposite directions. In such cases, quantities that indicate a direction of motion such as eq. (\ref{eq:velocity}) deliver different results than those that reflect the speed of particles, that is, the magnitude of the velocity. Furthermore, the assignment of a local velocity to a wave function, as in eq. (\ref{eq:velocity}), goes beyond the conventional interpretation of quantum mechanical wave functions and may require additional physical justification. 

 A particularly interesting case for the discussion of particle velocities are evanescent phenomena at potential steps or barriers. When particles encounter a (fully) reflective potential step, the wave function decays rapidly with a characteristic decay length depending on the mismatch between kinetic and potential energy. This decay, however, occurs \emph{within} the high-potential region, which distinguishes the quantum mechanical mode of motion from the classical one. The qualitative difference arises from the fact that quantum mechanical wave functions can have domains of negative kinetic energy, that is, regions where the local kinetic energy normalized to the particle density $(\psi(\mathbf{x})^{-1} \hat{\mathbf{p}}^2 \psi(\mathbf{x}))/2m$ takes negative values \cite{aharonov1992measurement,PhysRevA.48.4084}. This allows the particles to migrate to the high-potential region without violating energy conservation. Since a state of negative (local) kinetic energy has no classical equivalent, the question naturally arises as to what kind of motion it represents and, in particular, whether a negative kinetic energy gives rise to a physically meaningful particle velocity. Clearly, this question should be closely related to the topic of tunneling times, see Ref. \cite{Hauge89,Landauer94,winful2006tunneling,Sainadh2020} for overviews. The concept of a particle velocity, however, is rarely considered in this context. And if so \cite{Hirschfelder74, jayannavar87, deMoura90, Spiller90, Cahay92, Leavens93}, the topic is often treated from the perspective of the Bohmian interpretation of quantum mechanics \cite{Bohm52}, in which velocities play a central role. This is instructive but maybe not a generally accepted approach (see discussion in Ref. \citenum{Holland1995} for an overview). We believe that the study of energy-velocity relationships in evanescent phenomena can make a useful contribution to the tunneling time debate, particularly in helping to separate the various physical effects that, taken together, make this problem complicated. Such a study can also lead to an overall better understanding of how motion is represented in quantum mechanical wavefunctions.
 
The problem that first has to be solved, however, is the definition of a particle velocity that can be applied to both classically allowed and classically forbidden motion. A well-known approach used in the tunneling time problem is the embedding of additional degrees of freedom in the system whose change can be interpreted as a time measurement (Larmor clock) \cite{Baz1966, Rybachenko67, buttiker83, ramos20, spierings2021observation, demir2022tunneling}. In our work, we follow a related approach by studying the motion of massive particles in a system of two coupled waveguide potentials, where the population transfer between the waveguides effectively acts as a clock and allows particle velocities to be defined and determined. Application of this approach to evanescent phenomena at a reflective step potential reveals an energy-velocity relationship for classically forbidden motion. Regions of gain and loss are shown to speed up the motion of particles. These results will highlight the role of density gradients in wave functions for the motion of particles.

In preparation, consider the time evolution of a quantum mechanical system in which two degenerate states are coupled to each other with the coupling constant $J_0$. If the probability amplitude is initially entirely concentrated in one of the states, it is well known that the population in the initially unoccupied state follows $|\sin(J_0t)|^2$ as a function of time $t$ and, consequently, for small times increases like $(|J_0| t)^2$. Now we transfer this behavior into a propagating geometry by considering a steady-state particle stream guided in a system of two coupled waveguides. If the population is concentrated in one of the waveguides at $x=0$, we expect the population of the unoccupied waveguide to rise like $|\sin(J_0\,x/v)|^2$ with a velocity $v$. For small distances $x$, this can be approximated by $(|J_0|\,x/v)^2$. Thus, provided that the coupling constant $J_0$ is known, one can infer the velocity by measuring the spatial population build-up in the initially unoccupied state. Note that this definition of a particle velocity is invariant under a rotation of $J_0$ in the complex plane. Both dispersive (real-valued) and dissipative (imaginary-valued) types of coupling will give the same result as long as $|J_0|$ is conserved.

\begin{figure}[]
\includegraphics[width=8.65cm]{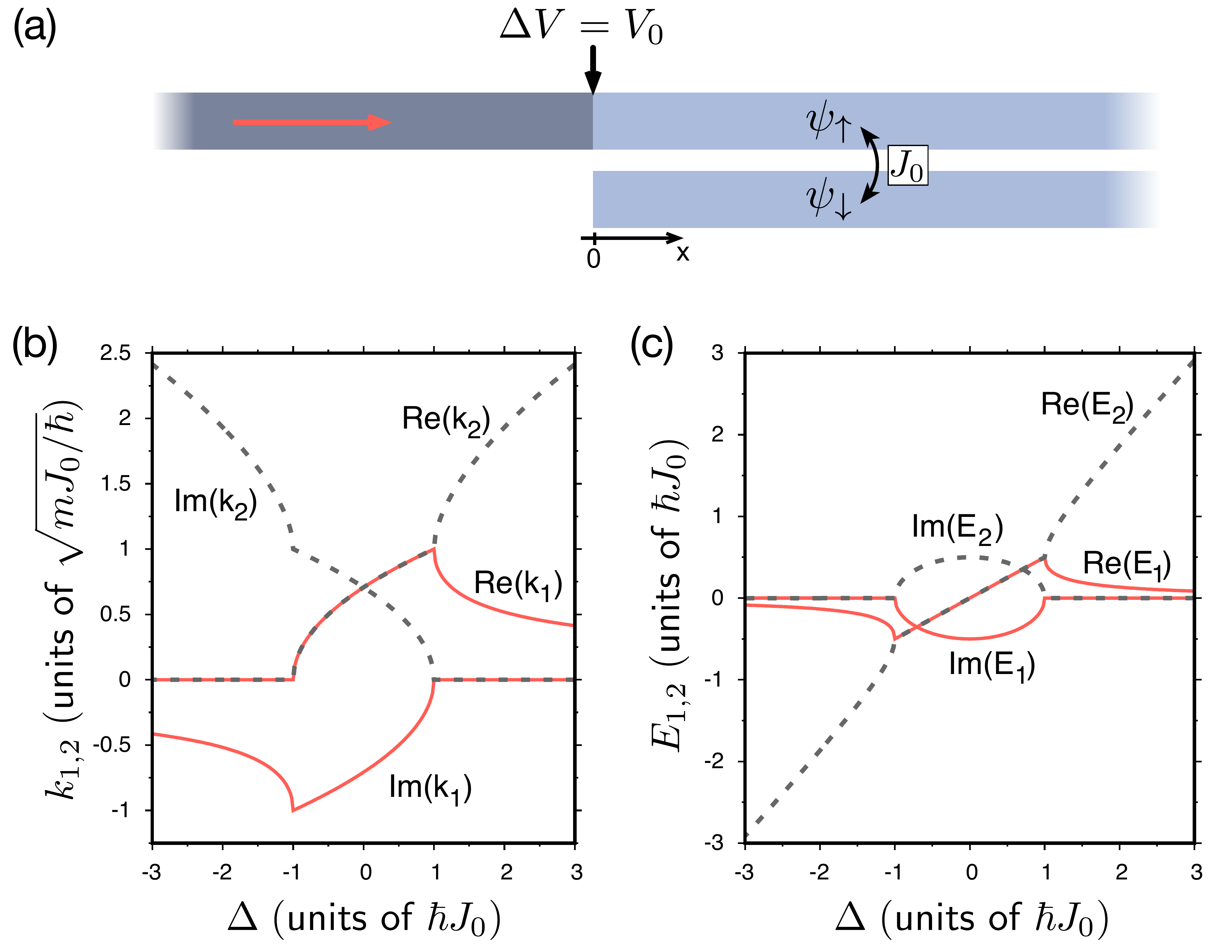}
\caption{\label{fig:coupled_waveguides} (a) A stream of particles (red arrow) is transversally confined in a waveguide potential and propagates towards a potential step (vertical black arrow). At the step, a second waveguide opens up. The particle transfer between the upper and the lower waveguide, described by the coupling constant $J_0>0$, effectively acts as a clock allowing particle velocities to be determined by considering the population build-up in $\psi_\downarrow$. (b) Solution of the coupled Schr\"odinger eqs. (\ref{schroedinger_up}) and (\ref{schroedinger_down}) in terms of the real and imaginary parts of $k_1$ (solid red lines) and $k_2$ (dashed gray lines) as a function of $\Delta=E+\hbar J_0-V_0$. (c) Real and imaginary part of $E_1=(\hbar k_1)^2/2m$ (solid red lines) and $E_2=(\hbar k_2)^2/2m$ (dashed gray lines).}
\end{figure}

More specifically, we consider a system where a stream of particles with mass $m$ is transversally confined in a waveguide potential and propagates towards a potential step at $x=0$ with $V(x)=0$ for $x<0$ and $V(x)=V_0$ for $x\ge 0$. At the position of the potential step, another waveguide potential opens up, which runs parallel to the first one, see Fig. 1a. It is assumed that the potential barrier between both waveguides is small enough that coupling between the wave functions in the upper ($\psi_\uparrow$) and lower ($\psi_\downarrow$) waveguide takes place and is quantitatively described by the coupling constant $J_0>0$. For $x\ge 0$, the steady state of the system is described by the coupled time-independent Schr\"odinger equations
\begin{eqnarray}
\label{schroedinger_up}
E \psi_\uparrow\;&&=\;-\frac{\hbar^{2}}{2m}\frac{\partial^2\psi_\uparrow}{\partial x^2}+V_0\psi_\uparrow+\hbar J_0 \left(\psi_\downarrow-\psi_\uparrow\right)\\ 
\label{schroedinger_down}
E \psi_\downarrow\;&&=\;-\frac{\hbar^{2}}{2m}\frac{\partial^2\psi_\downarrow}{\partial x^2}+V_0\psi_\downarrow+\hbar J_0 \left(\psi_\uparrow-\psi_\downarrow\right)\;. 
\end{eqnarray}
The total energy $E>0$ corresponds to the kinetic energy of the particles before they hit the potential step. Furthermore, we define the energy mismatch $\Delta=E+\hbar J_0-V_0$. 

\begin{figure*}
\includegraphics[width=18cm]{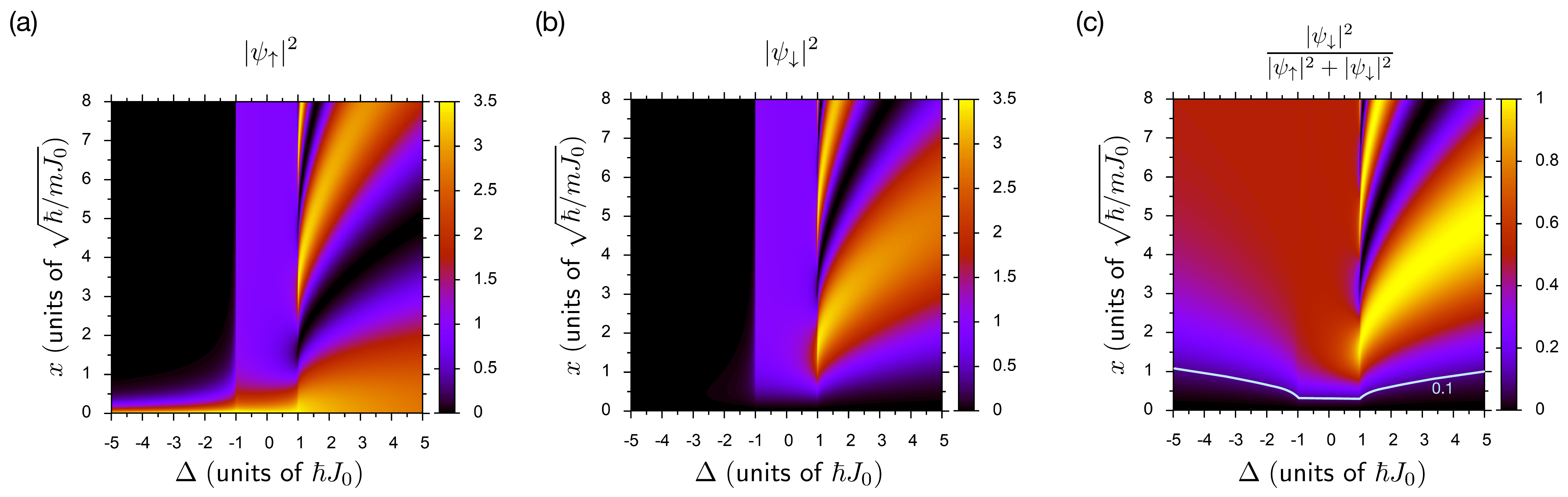}
\caption{\label{fig:full_solution} Particle densities $|\psi_{\uparrow,\downarrow}|^2$ as a function of energy mismatch $\Delta$ and position $x>0$ in the (a) upper and (b) lower waveguide. (c) Relative population in the lower waveguide $p_\downarrow=|\psi_\downarrow|^2/(|\psi_{\uparrow}|^2+|\psi_{\downarrow}|^2)$. Close to the potential step at $x=0$, both the classically allowed and the classically forbidden sides show the same relative population dynamics. The solid white line indicates a relative population of $p_\downarrow=0.1$.}
\end{figure*}

For $E>V_0$, classical propagation beyond the potential step at $x=0$ is possible and one expects that the wave functions in both waveguides can be represented by propagating plane waves whose amplitudes are harmonically modulated due to the waveguide coupling. A suitable ansatz to solving eqs. (\ref{schroedinger_up}) and (\ref{schroedinger_down}) is therefore $\psi_\uparrow \propto \cos(k_1 x)\exp(ik_2 x)$ and $\psi_\downarrow \propto \sin(k_1 x)\exp(ik_2 x)$. Note that the relative population of the waveguides $p_i=|\psi_i|^2/(|\psi_{\uparrow}|^2+|\psi_{\downarrow}|^2)$ with $i=\,\uparrow,\downarrow$ depends exclusively on the population transfer factors $\cos(k_1 x)$ and $\sin(k_1 x)$. It turns out that this ansatz works not only for $E>V_0$ but also for arbitrary energies. This general solution is uniquely determined by setting several requirements, which includes the assumption of an incoming wave with unity amplitude, the continuity of the wave functions $\psi_{\uparrow,\downarrow}$ and their derivatives $\partial_x\psi_{\uparrow,\downarrow}$, and the requirement that $\text{Re}(k_2)$ [$\text{Im}(k_2)$] is a positive, continuous, and monotonically increasing [decreasing] function of $\Delta$, as this reflects the expected behavior for $E \gg V_0$ [$E \ll V_0$]. With these requirements, the solution is 
\begin{eqnarray}
\label{psi_up}
\psi_\uparrow\;&&=\;\frac{2k_0}{k_0+k_2}\cos(k_1 x)\,\text{e}^{i k_2 x}\\ 
\label{psi_down}
\psi_\downarrow\;&&=\;-\frac{2 i k_0}{k_0+k_2}\sin(k_1 x)\,\text{e}^{i k_2 x}\;, 
\end{eqnarray}
with wavenumbers $k_{0,1,2}$ given by
\begin{eqnarray}
\label{k0}
k_0\;&&=\;\sqrt{2 m E}/\hbar\\ 
\label{k1}
k_1\;&&=\;m J_0/\hbar k_2\\ 
\label{k2}
k_2\;&&=\;\hbar^{-1}\sqrt{m\,\left( \Delta\pm\sqrt{\Delta^2-(\hbar J_0)^2} \right )}\;, 
\end{eqnarray}
where the '+' solution applies for $\Delta/\hbar J_0 > -1$ and the '-' solution otherwise. A graphical representation of this is shown in Fig. \ref{fig:coupled_waveguides}b and reveals three different regimes of propagation. For $\Delta/\hbar J_0>1$, $k_2$ is real-valued indicating classical propagation. For $\Delta/\hbar J_0<-1$, $k_2$ is imaginary indicating exponential decay of the wave function. For $|\Delta|/\hbar J_0 \le 1$, $k_{1,2}$ are complex conjugated to each other. In all three cases, the total energy is given by
\begin{equation}
\label{eq:energy}
E=\frac{(\hbar k_1)^2}{2m}+\frac{(\hbar k_2)^2}{2m} -\hbar J_0+V_0\;.
\end{equation}

To begin the discussion of particle velocities, consider the limit $\Delta/\hbar J_0 \rightarrow \infty$, which describes large excess of kinetic energy with respect to the energy scale of the coupling. In this case, Fig. \ref{fig:coupled_waveguides}c suggests that the energy associated with the population transfer between the waveguides $E_1=(\hbar k_1)^2/2m$ is negligible compared to $\Delta$. Accordingly, we expect that the energy-velocity relation of the particles within a waveguide is not affected by the presence of the neighboring waveguide and should be simply given by $v=\sqrt{2\Delta/m}$. There are two ways to recover this velocity from the solution of the coupled Schr\"odinger equations. On the one hand, the plane wave factor $\text{e}^{i k_2 x}$ in the wave functions with $k_2>0$ allows a direct assignment of a (group) velocity of $v=\hbar k_2/m$. With $k_2=\sqrt{2m\Delta}/\hbar$ following from eq. (\ref{k2}) in the limit $\Delta/\hbar J_0 \rightarrow \infty$, this gives the expected result for $v$. On the other hand, the population transfer factors $\cos(k_1 x)$ and $\sin(k_1 x)$ in the wave functions define a velocity via $k_1 x=J_0\:x/v$, or $v=J_0/k_1$, associated with the idea that the higher the velocity of the particles, the further away from the potential step the population build-up in the lower waveguide occurs. With $k_1=\sqrt{m/2 \Delta}\,J_0$ following from eq. (\ref{k1}) in the aforementioned limit, we equally recover $v=\sqrt{2\Delta/m}$. We note that the quantity determined in this way only reflects the magnitude of a velocity, since the population dynamics between the waveguides do not define a direction of motion. This can only be determined from the further physical context.

Now consider a strongly negative energy mismatch $\Delta/\hbar J_0 \rightarrow -\infty$. Also in this limit $|E_1| \ll |\Delta|$ holds, see Fig. \ref{fig:coupled_waveguides}c, and we do not expect the energy-velocity relation of the particles to be altered by the presence of the second waveguide. Since $k_2$ is purely imaginary, the plane wave factor $\text{e}^{i k_2 x}=\text{e}^{-|k_2| x}$ now describes an exponential decay, which does not provide a physically obvious interpretation in terms of a velocity or speed. However, the latter does not apply to the population transfer factors. For $\Delta/\hbar J_0 \rightarrow -\infty$, we find $k_1=\sqrt{m/2 \Delta}\,J_0$ just as before. However, this time we have $\Delta<0$, so that $k_1=\sqrt{m/2 |\Delta|}\,iJ_0$ becomes imaginary. Physically, this means that the coupling of the waveguides effectively switches from dispersive ($J_0>0$) to dissipative ($i J_0$) with the magnitude of the coupling remaining constant. As a result of this effectively dissipative coupling, the oscillating exchange of populations between the waveguides is replaced by a relaxation to equal occupation for $x\rightarrow \infty$, which can be seen from
\begin{equation}
\label{eq:relative_population}
p_\downarrow=\frac{\sinh^2(|k_1|x)}{\cosh^2(|k_1|x)+\sinh^2(|k_1|x)}\rightarrow\frac{1}{2}\;\;(\text{for}\; x\rightarrow \infty).
\end{equation}
This population transfer dynamics, however, still suggests the presence of a well-defined particle speed via $|k_1| x=J_0\:x/v$, or $v=J_0/|k_1|$, with the same physical interpretation as before: the higher the speed of the particles, the further away from the potential step the population build-up in the lower waveguide occurs. This gives a clear physical meaning to the notion of a speed for classically forbidden motion. Evaluating this expression, we obtain $v=\sqrt{2|\Delta|/m}$, which in this form covers both the classically allowed and classically forbidden case, as long as $|\Delta|/\hbar J_0 \rightarrow \infty$. We consider particles following this energy-velocity relation to be the most natural explanation to account for the fact that the population build-up in the lower waveguide depends on the energy of the incident particles - even in the classically forbidden case. We furthermore note that this expression is closely related to the B\"uttiker-Landauer time, which is recovered by setting $\tau=b/v$ with $b$ as the barrier width \cite{buttiker82}.

\begin{figure}[]
\includegraphics[width=8.0cm]{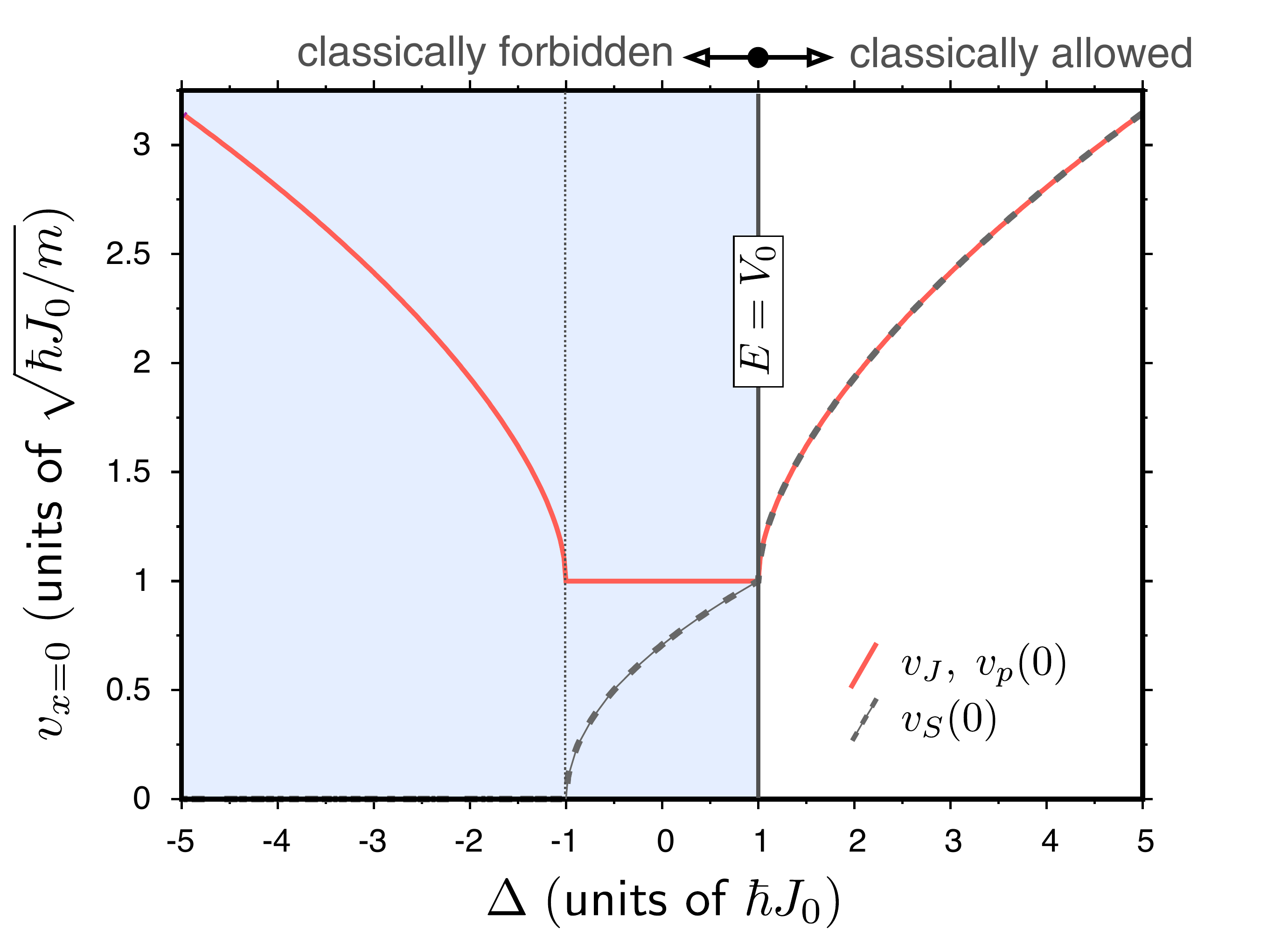}
\caption{\label{fig:velocity} Particle velocity at the potential step. The velocity $v_{J}$ [solid red line] is derived from the population transfer between the waveguides following eq. (\ref{eq:velocity2}). The velocity $v_p(0)$ [also solid red line] is derived from the local momentum associated with the wave function $\psi_\uparrow$ at $x=0$, as defined in the main text. The velocity $v_S(0)$ [dashed gray line] follows from eq. (\ref{eq:velocity}) and vanishes for $\Delta/\hbar J_0\le -1$. The distinction between classically allowed and forbidden (top of figure) is derived from the behavior of the (local) kinetic energy $T$, which is fed both by the motion along the waveguides and by the motion (hopping) between the waveguides, and accordingly is given by the first three terms in eq. (\ref{eq:energy}): $T=E_1+E_2-\hbar J_0$. Based on Fig. \ref{fig:coupled_waveguides}c, $T$ changes sign at $\Delta/\hbar J_0=1$ or equivalently $E=V_0$ [because of $\text{Re}(E_{1,2})=\hbar J_0/2$ and $\text{Im}(E_1)=-\text{Im}(E_2)$] and remains negative for all $\Delta<\hbar J_0$. As discussed in the introduction, a negative (local) kinetic energy has no classical analogue and can therefore serve as an indicator showing the transition between classically allowed and classically forbidden behavior.}
\end{figure}

We now consider the full solution of the model, which is represented in Fig. \ref{fig:full_solution}. Figures \ref{fig:full_solution}a,b show the population in the upper and lower waveguides for different energies $\Delta$ and positions $x>0$. The relative population in the lower waveguide $p_\downarrow$ is shown in Fig. \ref{fig:full_solution}c and generally indicates an oscillatory population transfer for positive and relaxation to equal populations for negative energy mismatch, which was already discussed in the limit $|\Delta|/\hbar J_0 \rightarrow \infty$. Close to the potential step $x/x_0 \ll 1$ with $x_0=\sqrt{\hbar/m J_0}$, however, the occupation of the lower waveguide $p_\downarrow$ is observed to become a mirror-symmetric function of the energy mismatch $\Delta$, which simply means that both the classically allowed and the classically forbidden sides show the same relative population dynamics. We now interpret this result with regard to particle speeds. Close to the potential step, a leading order approximation of the relative population build-up in the lower waveguide gives $p_\downarrow\simeq(|k_1|\,x)^2$ for all $\Delta$. A comparison with $p_\downarrow\simeq(J_0\,x/v)^2$ suggests $v=J_0/|k_1|$, or
\begin{equation}
\label{eq:velocity2}
v_{J}=\sqrt{\frac{|\Delta\pm\sqrt{\Delta^2-(\hbar J_0)^2}|}{m}}\;,
\end{equation}
where the '+' solution applies for $\Delta/\hbar J_0 > -1$ and the '-' solution otherwise. Note that the previously discussed result for vanishing coupling, that is $v=\sqrt{2|\Delta|/m}$, is reproduced by this relation. The obtained particle speed as a function of $\Delta$ is graphically represented in Fig. \ref{fig:velocity}. In particular, $v_J$ is observed to be mirror-symmetric with respect to $\Delta=0$ and to have a constant value of $v_{J,\text{min}}=\sqrt{\hbar J_0/m}$ in the range $|\Delta|/\hbar J_0 \le 1$. The latter indicates that in cases where the coupling energy becomes comparable to the kinetic energy, tunneling across the waveguides can have a significant impact on the motion of the particles along the waveguides. For comparison, we consider the phase gradient velocity $v_S$ following from eq. (\ref{eq:velocity}), which is equivalently given by $v_S=j/|\psi|^2$ with the probability current $j=\hbar(\psi^*\partial_x\psi- \text{h.c.})/2im$. $v_S(0)$ vanishes for $\Delta/\hbar J_0\le -1$, which is a reasonable result given that $v_S$ is a directional quantity and the step potential is fully reflective in this parameter range. However, as is demonstrated by the inter-waveguide population dynamics the vanishing of the phase gradient velocity should not be understood as absence of motion. 

As further comparison in Fig. \ref{fig:velocity}, we consider the particle speed $v_p$ derived from the local momentum associated with a wave function $\psi$ following $v_p=|\psi^{-1}\hat{p}\psi|/m$.
Evaluating $v_p$ for $\psi_\uparrow$ at $x=0$ indeed recovers the speed derived from the population transfer, see Fig. \ref{fig:velocity}. Similar to $v_J$, $v_p(0)$ does not indicate a direction. What is furthermore evident from the definition of $v_p$ (and the mirror symmetry of $v_J$ with respect to $\Delta=0$), is the fact that phase gradients and density gradients play complementary roles in indicating the motion of particles. While phase gradients indicate the particle speed of classically allowed propagation, this task is taken over by density gradients in the classically forbidden case.

\begin{figure}[]
\includegraphics[width=8.65cm]{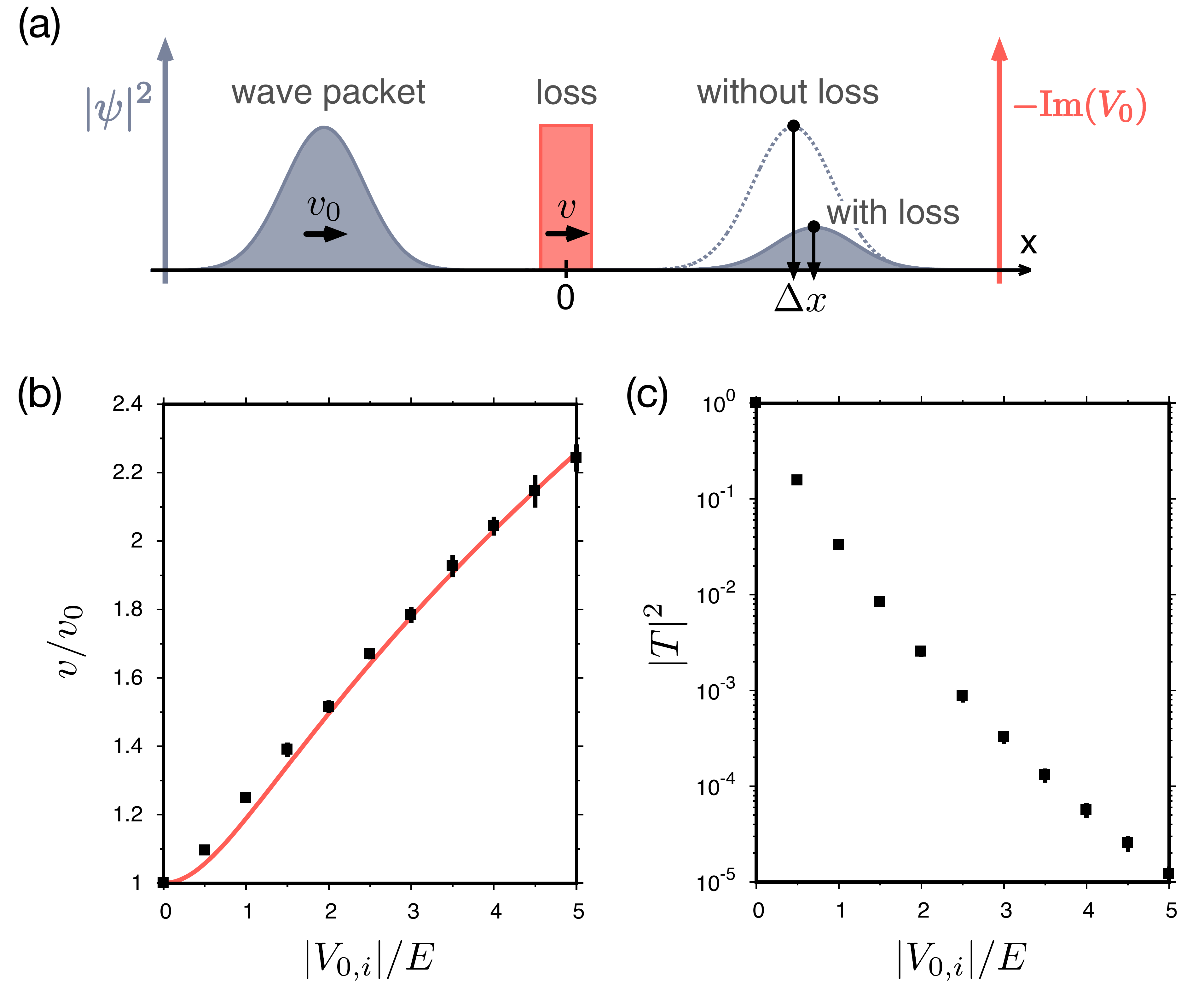}
\caption{\label{fig:wavepacket_velocity} (a) Scheme to investigate the velocity of particles passing a loss region. (b) Particle velocity in the region of the imaginary potential barrier as a function of the normalized barrier height $|V_{0,i}|/E$. The solid line follows from eq. (\ref{eq:velocity3}), the data points (squares) follow from numerically simulating the propagation of wave packets across the loss region. More specifically, we numerically integrate the time-dependent Schr\"odinger equation assuming narrow-bandwidth Gaussian wave packets with a fixed initial velocity and varying loss rates in the barrier ($V_{0,i}<0$). The position of the wave packets is determined by following their center of mass as a function of time. The particle velocity in the potential barrier $v$ is determined from position differences $\Delta x$ for various $V_{0,i}$ at the end of the simulation.  For the chosen simulation parameters (see below), we find a dispersion-related broadening of the wavepacket at the end of our simulations of approximately 2.5\% (for all $V_{0,i}$), which does not limit the ability to accurately determine these positions. The numerical integration is performed using the 4th-order Runge-Kutta method with adaptive time steps. Error bars (standard error of the mean) are derived from varying the spatial discretization in our integration scheme. Parameters used in the simulation: $m=6.5\cdot 10^{-36}\,\text{kg}$, $E=0.2\,\text{meV}$, $b=10\,\text{\textmu m}$ (barrier width), $\sigma=150\,\text{\textmu m}$ (standard deviation of initial Gaussian envelope). The parameters are chosen with regard to experimental realizations in low-dimensional photon or polariton gases. (c) Transmission through the loss region. The system parameters are the same as in (b).}
\end{figure}

So far we have considered $\Delta=E+\hbar J_0-V_0$ as real-valued quantity, which basically assumes real-valued potentials $V_0$. It is, however, desirable to extend the discussion to complex-valued $\Delta$ and thus complex potentials. Complex potentials describe scenarios with gain and loss and have been studied extensively in quantum mechanics \cite{molinas1996electron,ahmed2001schrodinger,muga2004complex,stutzle2005observation,guo2009observation}, also with regard to the tunneling time debate \cite{golub1990modest,muga1992equivalence,kovcinac2008tunneling}. It turns out that all results discussed so far do indeed remain valid in the case of complex $\Delta$. This is mainly due to the fact that the leading order approximation $p_\downarrow\simeq(|k_1|\,x)^2$ on which our speed measure $v=J_0/|k_1|$ is based still applies. 

In particular, the relation $v=\sqrt{2|\Delta|/m}$ for vanishing coupling remains valid in the case of complex $\Delta$. Rather than just stating this result, we also want to independently verify it. For this, we consider the case where the potential is purely imaginary $V_0=i\,V_{0,i}$. Then, it follows immediately that
\begin{equation}
\label{eq:velocity3}
\frac{v}{v_0}=\left ( 1+ \left ( \frac{V_{0,i}}{E} \right )^2 \right )^{1/4}\;,
\end{equation}
with $v_0=\sqrt{2E/m}$ as the velocity of freely propagating particles. This implies that both gain and loss increase the velocity of the particle stream. In Fig. \ref{fig:wavepacket_velocity}b, we show $v_0/v$ as a function of $|V_{0,i}|/E$ (solid line). For comparison, results derived from numerical simulations are included (points). These simulations study the propagation of wave packets across a loss region (imaginary potential barrier), see Fig. \ref{fig:wavepacket_velocity}a. Further details are given in the caption of that figure. Obviously, these two scenarios differ from each other in at least two aspects. First, they differ in the potential used. A potential barrier can produce resonator-like effects leading to deviating behavior compared to the step potential \cite{winful2006tunneling}. Second, the solution for the step potential derives from an energy eigenvalue problem, while in the case of the barrier wave packets with a finite energy uncertainty are used. Despite these differences, however, there is surprisingly high agreement between these two results. This is especially true when the potential becomes opaque, compare Fig. \ref{fig:wavepacket_velocity}b with Fig. \ref{fig:wavepacket_velocity}c. This agreement shows in particular that the observed speed up of the particles is not caused by a (frequency) filter effect, since such an effect is clearly absent in the energy eigenvalue problem. 

In this work, we introduce a measure for the speed of particles based on the population dynamics in a coupled waveguide system. This allows us to identify particle motion in evanescent phenomena, including both real and imaginary potentials. Phase and density gradients in quantum mechanical wave functions are found to play complementary roles in indicating particle motion. This result is in some contradiction with the Bohmian interpretation of quantum mechanics. Here, eq. (\ref{eq:velocity}) assumes the role of the guiding equation for particles and thus identifies phase gradients as the only signature of motion. The inter-waveguide population dynamics investigated in this work, however, suggests quite clearly that vanishing phase gradients do not imply the absence of motion. This ultimately suggests that the trajectories considered in Bohm's theory do not generally represent the actual motion of particles, but involve some level of statistical abstraction in certain situations. Last but not least, an essential motivation for our paper is to promote and prepare experimental work along the lines discussed. The scheme presented in this work has been chosen in particular with a view to possible experimental realizations. More specifically, we propose two-dimensional photon or exciton-polariton gases in optical microresonators as suitable experimental platforms \cite{bloch2022}, where precise control of the potential landscape \cite{kurtscheid20,Vretenar21,vretenar22}, including gain and loss, is possible. In such systems, photons or polaritons can be considered as massive particles propagating in the transverse plane of the resonator. Due to a small (and controlled) amount of leakage from the resonator, particle densities in the resonator plane can be fully reconstructed, regardless of whether they represent evanescent or propagating solutions. Performing a 'tunneling' experiment in such a system is therefore not dependent on the actual existence of a measurable barrier transmission, eliminating the need to work with barriers of finite width. In addition to implementations in optical systems, implementations in electronic and cold atom systems seem desirable \cite{thomas1999controlled}.

\begin{acknowledgments}
We thank Charlie Mattschas and Marius Puplauskis for carefully proofreading the manuscript and useful discussions. This work has received funding from the European Research Council (ERC) under the European Union’s Horizon 2020 research and innovation programme (Grant Agreement No. 101001512) and from the NWO (grant no. OCENW.KLEIN.453).
\end{acknowledgments}

\providecommand{\noopsort}[1]{}\providecommand{\singleletter}[1]{#1}%

\end{document}